\documentclass[12pt]{article}
\usepackage[totalwidth=470truept,totalheight=600truept]{geometry}
\usepackage{amsfonts,latexsym,amssymb,amsmath,graphicx,accents,eucal,slashed,subfigure}
\usepackage[T1]{fontenc}
\usepackage[hidelinks]{hyperref}

\def\theequation{\arabic{section}.\arabic{equation}}

\renewcommand{\theequation}{\thesection.\arabic{equation}}
\global\arraycolsep=1truept

\numberwithin{equation}{section}
\renewcommand{\theequation}{\arabic{section}.\arabic{equation}}

\begin{document}

\bigskip \phantom{C}

\vskip 1.5truecm

\begin{center}
{\huge \textbf{Inconsistency Of Minkowski}}

\vskip.4truecm

{\huge \textbf{Higher-Derivative Theories}}

\vskip1truecm

\textsl{Ugo G. Aglietti}

\vskip .1truecm

\textit{Dipartimento di Fisica, Universit\`{a} di Roma \textquotedblleft La
Sapienza\textquotedblright , }

\textit{Piazza Aldo Moro 2, 00185 Rome, Italy}

ugo.aglietti@roma1.infn.it

\vskip.7truecm

\textsl{Damiano Anselmi}

\vskip .1truecm

\textit{Dipartimento di Fisica ``Enrico Fermi'', Universit\`{a} di Pisa, }

\textit{Largo B. Pontecorvo 3, 56127 Pisa, Italy}

\textit{and INFN, Sezione di Pisa,}

\textit{Largo B. Pontecorvo 3, 56127 Pisa, Italy}

damiano.anselmi@unipi.it

\vskip1truecm

\textbf{Abstract}
\end{center}

We show that Minkowski higher-derivative quantum field theories are
generically inconsistent, because they generate nonlocal, non-Hermitian
ultraviolet divergences, which cannot be removed by means of standard
renormalization procedures. By \textquotedblleft Minkowski\
theories\textquotedblright\ we mean theories that are defined directly in
Minkowski spacetime. The problems occur when the propagators have complex
poles, so that the correlation functions cannot be obtained as the analytic
continuations of their Euclidean versions. The usual power counting rules
fail and are replaced by much weaker ones. Self-energies generate complex
divergences proportional to inverse powers of D'Alembertians. Three-point
functions give more involved nonlocal divergences, which couple to infrared
effects. We illustrate the violations of the locality and Hermiticity of
counterterms in scalar models and higher-derivative gravity.

\vfill\eject

\newpage

\section{Introduction}

\label{s0} \setcounter{equation}{0}

The ultraviolet structure of quantum field theories is notoriously a
fundamental problem in high-energy physics. Nowadays, with the Large Hadron
Collider currently running at a center-of-mass energy of 13TeV, the standard
model is experimentally verified at the TeV scale. On the theoretical side,
the renormalizability of the standard model, together with the relatively
small value found for the Higgs mass, $m_{H}\simeq 125$GeV, imply that the
model could be valid at energies much higher than the ones investigated so
far. The interest in a high-energy modification of the standard model is,
therefore, rather limited, on the practical side. As far as quantum gravity
is concerned, the situation is different. The negative mass dimension of the
coupling constant (compared with the dimensionless gauge couplings of the
standard model) makes the Hilbert--Einstein action nonrenormalizable \cite%
{thooftveltman}. This fact, together with the difficulties to build up a
phenomenology, render the investigation of alternative high-energy
structures of quantum gravity more attractive.

An interesting class of higher-derivative quantum field theories are those
whose propagators have complex poles. In that case, the Euclidean and
Minkowski versions of the theories are not related to each other by the
analytic continuation. In this paper, we concentrate on the Minkowski
formulation of such theories, that is to say we integrate the loop energies
along the real axis. We show that such theories are generically
inconsistent, because they violate both the locality and Hermiticity of
counterterms. For example, the one-loop bubble diagram $\Sigma (p)$ of
massless higher-derivative scalar fields in six spacetime dimensions
evaluates to 
\begin{equation}
\Sigma (p)=-\frac{\,M^{4}}{2(4\pi )^{3}}\left[ \frac{M^{2}}{(p^{2})^{2}}-%
\frac{i}{p^{2}}\right] \ln \left( \frac{\Lambda _{UV}^{2}}{M^{2}}\right)
+\cdots ,  \label{sketch}
\end{equation}%
where $\Lambda _{UV}$ is a hard ultraviolet cutoff on the space momenta, $M$%
\ is the scale associated with the higher-derivative terms and the dots
denote convergent terms. Similar results occur in four dimensions, when
vertices carry derivatives. More involved nonlocal structures appear in
triangle diagrams. Moreover, gauge symmetries are unable to protect the
locality and Hermiticity of counterterms. We prove this fact by extending
the calculations to a model of higher-derivative quantum gravity \cite%
{stelle}.

The rules of power counting obeyed by Minkowski higher-derivative theories
are much weaker than the standard ones, because a propagator calculated on
the pole of another propagator falls off half less rapidly than expected.
This property implies that the higher-derivative terms often have an
\textquotedblleft antiregulating\textquotedblright\ effect, in the sense
that they enhance divergences rather than suppressing them. The divergences (%
\ref{sketch}) can also be related to specific pinch singularities occurring
for $p^{2}\rightarrow 0$, which have no direct analog in standard field
theories.

It is well known that, in general (for example, when the free propagators
have poles infinitesimally close to the real axis, in the second and fourth
quadrants), Minkowski higher-derivative theories are physically
unacceptable, because they violate perturbative unitarity. Our results show
that when the free propagators also contain poles that are located at finite
distances from the real axis, in the first and third quadrants, the theories
are in general unacceptable from the mathematical point of view, because
they violate both the locality and the Hermiticity of counterterms.

The problems we have found do not occur in Euclidean theories, Lee--Wick
models \cite{leewick} or Minkowski theories that are analytically equivalent
to their Euclidean versions. In those cases, nonlocal divergences may appear
in some intermediate steps of the calculations (examples being the residues
of the integrals on the energies, if taken separately), but they cancel out
at the end.

The paper is organized as follows. In sect. \ref{secprop} we study some key
aspects of the higher-derivative Minkowski propagator. In sect. \ref{bubbles}%
, we calculate the nonlocal divergent part of the bubble diagram in six
dimensions and generalize the calculation to the bubble diagram with
nontrivial numerators in four dimensions. In sect. \ref{triangles}, we study
the one-loop triangle diagram and provide an interpretation for its nonlocal
divergent part. In sect. \ref{pc} we investigate the modified power counting
of Minkowski quantum field theories. In sect. \ref{QG}, we study a
higher-derivative version of quantum gravity in four dimensions and prove
that gauge symmetries fail to protect the locality and Hermiticity of
counterterms. Finally, in sect. \ref{concl} we draw our conclusions.

In appendix \ref{tricks} we show that the dimensional regularization (of the
integrals on the space momenta) allows us to apply the residue theorem on
the energy integrals, even when they are divergent. In appendix \ref{appe}
we discuss the gauge fixing of Minkowski higher-derivative gravity and show
that the Ward--Takahashi--Slavnov--Taylor (WTST) identities \cite{WTST} are
also plagued with nonlocal divergences.

\section{Higher-derivative propagator}

\label{secprop} \setcounter{equation}{0}

The standard propagator of a spinless particle reads 
\begin{equation}
\Delta (p,m)=\frac{1}{p^{2}-m^{2}+i\epsilon }.  \label{stap}
\end{equation}%
To increase the convergence of the loop integrals for large virtualities, $%
|p^{2}|\gg m^{2}$, it is natural to introduce additional powers of $p^{2}$
in the denominator to obtain, for example, a modified propagator of the form 
\begin{equation}
S(p,m)=\frac{1}{p^{2}-m^{2}+i\epsilon }\,\frac{M^{4}}{(p^{2})^{2}+M^{4}}.
\label{modified}
\end{equation}%
For small virtualities, $|p^{2}|\ll M^{2}$, the modified propagator
approaches the standard one, $S(p,m)\simeq \Delta (p,m)$, while in the
asymptotic region 
\begin{equation}
|p^{2}|\gg M^{2},  \label{as_reg}
\end{equation}%
it decays quite faster:%
\begin{equation*}
S(p,m)\simeq \frac{M^{4}}{(p^{2})^{3}}.
\end{equation*}%
By means of a partial fractioning in $p^{2}$, the higher-derivative
propagator (\ref{modified}) can also be written as 
\begin{eqnarray}
S(p,m) &=&\frac{M^{4}}{M^{4}+m^{4}}\frac{1}{2\omega _{\epsilon }(p_{s})}%
\left[ \frac{1}{p_{0}-\omega _{\epsilon }(p_{s})}-\frac{1}{p_{0}+\omega
_{\epsilon }(p_{s})}\right]  \notag \\
&&-\frac{M^{2}}{M^{2}-im^{2}}\frac{1}{4\Omega (p_{s})}\left[ \frac{1}{%
p_{0}-\Omega (p_{s})}-\frac{1}{p_{0}+\Omega (p_{s})}\right] +  \notag \\
&&\,\,\,\,\,\,\,\,\,\,\,\,\,\,\,\,\,\,\,-\frac{M^{2}}{M^{2}+im^{2}}\frac{1}{4%
\bar{\Omega}(p_{s})}\left[ \frac{1}{p_{0}-\bar{\Omega}(p_{s})}-\frac{1}{%
p_{0}+\bar{\Omega}(p_{s})}\right] ,  \label{partf}
\end{eqnarray}%
where $\omega _{\epsilon }(p_{s})\equiv \sqrt{p_{s}^{2}+m^{2}-i\epsilon }$, $%
\Omega \mathcal{(}p_{s})\equiv \sqrt{p_{s}^{2}-iM^{2}}$, $p^{\mu }=\left(
p_{0},\mathbf{p}\right) $ and $p_{s}=|\mathbf{p}|$. The poles are located at 
$p_{0}=\pm \omega _{\epsilon }(p_{s})$, $p_{0}=\pm \Omega (p_{s})$ and $%
p_{0}=\pm \bar{\Omega}(p_{s})$, where the bar denotes the complex
conjugation. Since poles are present in every quadrant, the Euclidean theory
and the Minkowski theory are not related in a simple way. In this paper we
concentrate on the Minkowski theory, that is to say we assume that the loop
integral is defined by integrating the energy along the real axis.

In general, the factor 
\begin{equation}
\frac{M^{4}}{(p^{2})^{2}+M^{4}}\leqslant 1  \label{factor}
\end{equation}%
is expected to have an ultraviolet regulating effect, by suppressing the
states with $\left\vert p^{2}\right\vert \gg M^{2}$. We show that it is not
the case in Minkowski theories. Actually, often the factor (\ref{factor})
roughly has an opposite, \textquotedblleft
anti-regulating\textquotedblright\ effect.

For large space momenta, $p_{s}\gg M$, the complex poles come close to the
real axis, since 
\begin{equation}
\Omega \mathcal{(}p_{s})\equiv \sqrt{p_{s}^{2}-iM^{2}}\simeq p_{s}-i\eta
,\qquad \bar{\Omega}(p_{s})\simeq p_{s}+i\eta ,  \notag
\end{equation}%
where $\eta $ is the small positive quantity 
\begin{equation*}
\eta =\eta \left( p_{s}\right) \equiv \frac{M^{2}}{2p_{s}}.
\end{equation*}%
Therefore, in the asymptotic region $p_{s}\gg M$ the terms involving $M$
effectively act as $\pm i\eta $ prescriptions for the propagation of exotic,
high-energy excitations on the light cone, with the large lifetimes 
\begin{equation*}
\tau =\tau \left( p_{s}\right) \approx \frac{p_{s}}{M^{2}}.
\end{equation*}%
Both $\epsilon $ and $\eta \left( p_{s}\right) $ are positive quantities,
but $\epsilon $ is infinitesimal, while $\eta \left( p_{s}\right) $ is small
and finite.

Since $\eta \left( p_{s}\right) \rightarrow 0$ for $p_{s}\rightarrow +\infty 
$, there is a pinch singularity of the pole located in the first quadrant
with the two poles located in the fourth quadrant, and a similar pinch
singularity of the pole located in the third quadrant with the two poles
located in the second quadrant. The violations of power counting that we
find in the next sections can be traced back to this pinching and ultimately
to the presence of both the $+i\eta $ and the $-i\eta $ terms at $p_{s}\gg M$%
.

\section{Bubble diagrams}

\label{bubbles} \setcounter{equation}{0}

In this section we compute the nonlocal divergent parts of the
higher-derivative, one-loop scalar bubble diagrams in six and four
dimensions, with trivial and nontrivial numerators.

As usual, the ultraviolet divergent part is a sum of powerlike divergences
and logarithmic divergences. The powerlike divergences are less interesting
than the logarithmic ones for the purpose of singling out inconsistencies,
because they depend on the subtraction scheme and can be removed in a
renormalization-group invariant way. The one-loop logarithmic divergences,
on the contrary, do not depend on the regularization scheme, and provide
meaningful tests of the locality of counterterms. For these reasons, we
focus our attention mostly on them. We either use the dimensional
regularization technique or a sharp cutoff $\Lambda _{UV}$ on the space
momenta of the loops, according to convenience. We always convert the
outcome to the cutoff notation.

\subsection{Bubble diagram in six dimensions}

\setcounter{equation}{0}\label{bubb}

The bubble diagram with different masses in $D$ spacetime dimensions gives
the loop integral 
\begin{equation}
\Sigma (p)=\int\limits_{k_{s}\leqslant \Lambda _{UV}}\frac{\mathrm{d}^{D-1}%
\mathbf{k}}{(2\pi )^{D-1}}\int\limits_{-\infty }^{+\infty }\frac{\mathrm{d}%
k_{0}}{2\pi }\hspace{0.01in}S\left( k,m_{1}\right) S\left( k-p,m_{2}\right) ,
\label{bubble}
\end{equation}%
where $\hspace{0.01in}S\left( p,m\right) $ is given in formula (\ref%
{modified}), $\Lambda _{UV}$ is an ultraviolet cutoff on the space momenta $%
\mathbf{k}$ and $k_{s}=|\mathbf{k}|$.

Since the higher-derivative theory is relativistically invariant, $\Sigma
(p) $ is expected to be a function of $p^{2}$ only\footnote{%
To be rigorous, one should use an ultraviolet regularization that preserves
Lorentz invariance, such as the dimensional regularization, instead of a
cutoff on the space momenta. However, we are only interested in the
logarithmic divergences, which, as already noted, are independent of this
choice.}. It is then convenient to consider a timelike external momentum, $%
p^{2}>0$, and select a Lorentz frame in which $p^{\mu }$ has only the time
component, $p^{\mu }=\left( p_{0},\mathbf{0}\right) $. Once the loop
integral is evaluated, we can retrieve the Lorentz invariant result by means
of the replacement $p_{0}^{2}\rightarrow p^{2}$. The values of the bubble
diagram for a spacelike external momentum, $p^{2}<0$, are obtained by means
of the analytic continuation in $p^{2}$.

The first step is to integrate $k_{0}$ over the real line, which we can make
in two equivalent ways. The first method is by applying the residue theorem,
after closing the integration path with a semicircle at infinity, say in the
upper half $k_{0}$ plane. The second method involves the partial fractioning
in $k_{0}$. Since there are only simple poles for $p_{0}\neq 0$, one ends up
with a $k_{0}$ integrand of the form 
\begin{equation*}
\sum_{i=1}^{12}\frac{c_{i}}{k_{0}-s_{i}},
\end{equation*}%
where $s_{i}$ are the poles of the propagators and $c_{i}$ are coefficients
that depend on the nonintegrated variables $k_{s},p_{0},m_{1},m_{2},M$.
Since each propagator contains six poles, there are twelve terms in total.
Then, one integrates over $k_{0}$ term by term. In practice, $1/\left(
k_{0}-s\right) $ gives $\pm i\pi $, depending on whether $\mathrm{Im}\,s>0$
or $\mathrm{Im}\,s<0$.

The next step is to integrate over the space momentum $\mathbf{k}$. The
angular integration is trivial, because of our choice of the Lorenz frame,
and gives the volume $\Omega _{D-2}$ of the unit sphere in $D-2$ dimensions.

Since we are only interested in the ultraviolet divergences for $%
k_{s}\rightarrow \infty $, we expand the integrand for large $k_{s}$, in
order to avoid special functions due to the $k_{s}$ integral. Consider the
residues calculated at the first step. In each of them, either $k^{2}$ or $%
(k-p)^{2}$ is equal to a constant. The two cases are symmetrical, so we just
assume $k^{2}=$ constant. Then, the propagator $S\left( k,m_{1}\right) $
gives a contribution $\sim 1/k_{s}$ for large $k_{s}$, by formula (\ref%
{partf}). Instead, the propagator $S\left( k-p,m_{2}\right) $ behaves as $%
1/((k-p)^{2})^{3}\sim 1/(p\cdot k)^{3}\sim 1/k_{s}^{3}$ (having used $p\cdot
k=p_{0}k_{0}$). The product of the two behaves as $1/k_{s}^{4}$, so the $%
k_{s}$ integral diverges like 
\begin{equation*}
\int^{\Lambda _{UV}}\frac{k_{s}^{D-2}\mathrm{d}k_{s}}{k_{s}^{4}}.
\end{equation*}%
However, it is easy to check that the contributions of such type coming from
the poles with $k^{2}=$constant compensate analogous contributions coming
from the poles with $(k-p)^{2}=$constant. In the end, the integrand of (\ref%
{bubble}) behaves as $1/k_{s}^{5}$, so we find that the leading $\Sigma $
divergence is proportional to 
\begin{equation*}
\int^{\Lambda _{UV}}\frac{\mathrm{d}k_{s}}{k_{s}^{7-D}}.
\end{equation*}%
We conclude that for $D<6$ the bubble diagram is ultraviolet finite, while
it is divergent for $D\geqslant 6$. In particular, at $D=6$ there is a
logarithmic divergence, 
\begin{equation*}
\,\,\,\int_{M}^{\Lambda _{UV}}\frac{\mathrm{d}k_{s}}{k_{s}}\,=\,\frac{1}{2}%
\ln \left( \frac{\Lambda _{UV}^{2}}{M^{2}}\right) ,
\end{equation*}%
which leads to the final result 
\begin{eqnarray}
\Sigma (p) &=&\,\frac{\,1}{12(4\pi )^{3}}\,\frac{M^{6}}{\left(
M^{2}+im_{1}^{2}\right) \left( M^{2}+im_{2}^{2}\right) }\left\{ \frac{1}{%
\left( p^{2}\right) ^{2}}\,\left[
m_{1}^{4}+m_{2}^{4}-4iM^{2}(m_{1}^{2}+m_{2}^{2})-6M^{4}\right] \right. + 
\notag \\
&+&\left. \frac{3i}{p^{2}}\,\left[ 2M^{2}+i(m_{1}^{2}+m_{2}^{2})\right]
\right\} \log \left( \frac{\Lambda _{UV}^{2}}{M^{2}}\right) \,+\,(\mathrm{%
finite}),  \label{assurdo}
\end{eqnarray}%
where by \textquotedblleft $\mathrm{finite}$\textquotedblright\ we mean
terms that are finite or infinitesimal for $\Lambda _{UV}\rightarrow +\infty 
$.

The divergent part is nonlocal, equal to the sum of a term proportional to $%
1/(p^{2})^{2}$ plus a term proportional to $1/p^{2}$. Differently from the
usual divergences of local theories, which are anti-Hermitian, the ones of (%
\ref{assurdo}) are not, since the coefficients have nontrivial real and
imaginary parts. For these reasons, we cannot absorb the divergent part in
the usual way, by shifting the bare masses, rescaling the bare fields and
adding new local, Hermitian terms to the Lagrangian. We cannot even add
nonlocal Hermitian terms. We conclude that the locality and Hermiticity of
counterterms are both violated.

Since we are exploring an uncharted territory, we wish to make an explicit
check of Lorentz invariance and analyticity. We consider the usual bubble in
the case $p^{2}<0$, by taking $p^{\mu }=(0,\mathbf{p}_{s})$. As in the
previous computation, we integrate over $k_{0}$ by means of the residue
theorem or the partial fractioning in $k_{0}$. Then we have to integrate
over the angles, which is a nontrivial operation now. Writing $k^{\mu
}=(k_{0},\mathbf{k}_{s})$, we switch to spherical coordinates, letting $%
\theta $ denote the angle between $\mathbf{p}_{s}$ and $\mathbf{k}_{s}$. The
integral over the remaining three angles is trivial, which leads to the
replacement 
\begin{equation*}
\mathrm{d}^{5}k_{s}\,\,\rightarrow \,\,2\pi ^{2}\mathrm{d}%
k_{s}k_{s}^{4}\left( 1-u^{2}\right) \mathrm{d}u,
\end{equation*}%
where $u\equiv \cos \theta $. At this point, we should expand the integrand
for large $k_{s}$. This cannot be done naively without generating $u$ poles.
For example, consider a typical denominator that is met in the calculation,
such as%
\begin{equation}
\frac{1}{2uk_{s}p_{s}-p_{s}^{2}-2iM^{2}}.  \label{rat}
\end{equation}%
If we expand it for large $k_{s}$, we obtain divergent $u$ integrals.
However, according to (\ref{rat}), the $u$ pole has a positive imaginary
part. If we first replace $u$ by $u-i\epsilon $, with $\epsilon $
arbitrarily small, then the expansion for large $k_{s}$ is safe. So doing,
no fictitious singularity is generated.

The procedure works as long as the integrand can be arranged so that the
powers $1/(u-i\epsilon )^{n}$ do not mix with the powers $1/(u+i\epsilon
)^{n}$. It can be shown that the bubble diagram has this property. Carrying
on the computation to the end, we find (\ref{assurdo}) again.

\subsection{Bubble diagram in four dimensions}

As shown in the previous section, the bubble diagram with unit numerator has
nonlocal divergences only in dimensions $D\geqslant 6$. On the other hand,
bubble diagrams with nontrivial numerators may have nonlocal divergences
also in four dimensions. In this section we study typical one-loop integrals
of this type. Their applications to higher-derivative gravity will be
considered in section \ref{QG}.

We assume that the propagator has again the form (\ref{modified}) and the
vertices contain an arbitrary number of derivatives. We study the scalar
integrals%
\begin{equation}
I_{r,n}(p)\equiv \int \frac{\mathrm{d}^{D}k}{(2\pi )^{D}}(k\cdot
p)^{r}(k^{2})^{n}S(k,m)S(p-k,m).  \label{imn}
\end{equation}%
The nonlocal divergent part can be calculated with the method explained in
the previous subsection. We set $p^{2}>0$ and choose $p^{\mu }=(p_{0},%
\mathbf{0})$. First, we integrate on the energy by means of the residue
theorem, closing the integration path on the upper half complex plane. In
appendix \ref{tricks} we show that, if we use the dimensional
regularization, the energy integral can always be evaluated by summing the
residues, even when it is divergent, because the contribution of the
integration path at infinity is always zero.

Then we remain with the integral on the space momentum $\mathbf{k}$. The
logarithmic nonlocal divergences $I_{r,n}^{\text{nl\hspace{0.01in}d}}$ of $%
I_{r,n}$ are obtained by expanding the integrand in powers of the absolute
value $k_{s}=|\mathbf{k}|$ and isolating the contributions proportional to $%
\mathrm{d}k_{s}/k_{s}$.

We report results in various cases, starting from the massless limit. There,
we find 
\begin{equation}
I_{r,n}^{\text{nl\hspace{0.01in}d}}=c_{r,n}\left( \frac{-iM^{2}}{2}\right)
^{r+n}\frac{M^{2}}{(4\pi )^{2}p^{2}}\log \left( \frac{\Lambda _{UV}^{2}}{%
M^{2}}\right) ,  \label{irn}
\end{equation}%
where $c_{r,n}$ are positive integer numbers. As anticipated, the pole $%
1/(4-D)$ of the dimensional regularization has been converted into the
logarithm $\log (\Lambda _{UV}/M)$ of a generic ultraviolet cutoff $\Lambda
_{UV}$ divided by $M$. The lowest-order coefficients are 
\begin{eqnarray}
c_{0,0} &=&0,\qquad c_{0,1}=1,\qquad c_{1,0}=0,\qquad c_{1,1}=3,\qquad
c_{1,2}=2,\qquad c_{2,1}=1,\qquad c_{2,2}=14,  \notag \\
c_{3,1} &=&15,\qquad c_{1,3}=12,\qquad c_{2,3}=4,\qquad c_{3,2}=2,\qquad
c_{3,3}=60.  \label{cc}
\end{eqnarray}

The basic features of these results may be justified by evaluating the
residues associated with the propagators of $I_{r,n}$, as explained in
subsection \ref{bubb}. The residues with $k^{2}=$constant have a superficial
degree of divergence $\omega _{1}=D-5+r$, where $D-1$ powers come from the $%
k_{s}$ integration measure, $-1$ and $-3$ from the propagators and $r$ from
the numerator. Instead, the residues with $\left( k-p\right) ^{2}=$constant
have degree of divergence $\omega _{2}=D-5+r+n$, where the $n$ additional
powers come from the numerator, using $k^{2}\sim 2k\cdot p$. For $r=1$, $n=0$%
, we have $\omega _{1}=\omega _{2}=D-4$, so ultraviolet logarithmic
divergences are expected from both types of residues in $D=4$. However,
formula (\ref{cc}) shows that the coefficient $c_{1,0}$ vanishes. This may
be interpreted as an eikonal cancellation between the two types of residues.
For $r=0$, $n=1$, we have $\omega _{2}>\omega _{1}=D-4$, so the cancellation
cannot occur in $D=4$. Indeed, $c_{0,1}$ is different from zero. More
generally, cancellations between the ultraviolet divergences of the residues
are unlikely to occur for $n>0$. Indeed, all the coefficients (\ref{cc})
with $n>0$ are nonvanishing.

In the massive case, we find formula (\ref{irn}) with coefficients $c_{r,n}$
that depend on $m/M$. The lowest-order ones are%
\begin{equation*}
c_{0,1}=\frac{M^{2}}{M^{2}+im^{2}},\qquad c_{1,1}=\frac{3M^{2}+im^{2}}{%
M^{2}+im^{2}},\qquad c_{1,2}=2\left( 1+\frac{im^{2}}{M^{2}}\right) .
\end{equation*}%
Instead, if we replace the propagators $S(p,m)$ with the more general ones%
\begin{equation*}
S(p,m,\mu )=\frac{1}{p^{2}-m^{2}+i\epsilon }\,\frac{M^{4}}{(p^{2}-\mu
^{2})^{2}+M^{4}},
\end{equation*}%
we obtain formula (\ref{irn}) with%
\begin{eqnarray*}
c_{0,1} &=&\frac{M^{2}}{M^{2}+i\Delta m^{2}},\qquad c_{1,1}=\frac{%
3M^{2}+i\Delta m^{2}}{M^{2}+i\Delta m^{2}}, \\
c_{1,2} &=&2\left( 1+\frac{i\Delta m^{2}}{M^{2}}+\frac{6i\mu ^{2}}{%
M^{2}+i\Delta m^{2}}-\frac{2\mu ^{2}\Delta m^{2}}{M^{2}(M^{2}+i\Delta m^{2})}%
\right) ,
\end{eqnarray*}%
where $\Delta m^{2}=m^{2}-\mu ^{2}$.

We see that the locality and Hermiticity of counterterms are violated again.
The \ nonlocal behavior is always of the form $1/p^{2}$, but it must be
recalled that the integrals (\ref{imn}) contain $r$ powers of $p^{\mu }$ in
the numerator, through the term $(k\cdot p)^{r}$. If we divide by those
powers, the true nonlocal behavior of the divergent part is $\sim
1/(p^{2})^{(2+r)/2}$.

\section{Triangle diagrams}

\label{triangles} \setcounter{equation}{0}

In this section we consider the one-loop three-point function. A peculiar,
double-logarithmic structure%
\begin{equation*}
\ln \left( \frac{Q^{2}}{\mu ^{2}}\right) \ln \left( \frac{\Lambda _{UV}^{2}}{%
M^{2}}\right)
\end{equation*}%
is found, coming from the overlap between the collinear (infrared) region $%
\mu ^{2}\ll k_{\perp }^{2}\ll Q^{2}$ and the ultraviolet region $M^{2}\ll
k_{s}^{2}\ll \Lambda _{UV}^{2}$, where $k_{\perp }$ is the transverse loop
momentum, $Q$ is the hard scale and $\mu $ is the virtuality of external
legs.

As in the case of the bubble diagram, in order to effectively generate such
divergences, we need to introduce nontrivial numerators, which we assume to
be scalar for simplicity. Specifically, we consider the following
three-index family of amplitudes: 
\begin{equation}
I_{r,n,t}\left( p_{1},p_{2}\right) =\int \frac{\mathrm{d}^{D}k}{(2\pi )^{D}}%
(k\cdot p_{1})^{r}(k^{2})^{n}(k\cdot p_{2})^{t}S(k,m)S\left(
k+p_{1},m\right) S\left( k+p_{2},m\right) .  \label{irnt}
\end{equation}%
The nonlocal divergent part $I_{r,n,t}^{\text{nl\hspace{0.01in}d}}$ can be
calculated with a procedure similar to the one used for the bubble diagram.
First, we use Lorentz invariance and analyticity to impose that the incoming
momenta are all timelike, i.e. 
\begin{equation}
p_{1}^{2}>0,\qquad p_{2}^{2}>0,\qquad (p_{1}-p_{2})^{2}>0,  \label{analit}
\end{equation}%
and choose $p_{1}^{\mu }=(E_{1},\mathbf{0})$, while $p_{2}^{\mu }=(E_{2},%
\mathbf{p}_{2})$ remains generic. Then we integrate on the energy $k_{0}$ by
means of the residue theorem or a partial fractioning. At that point, we
expand the integrand in powers of $1/k_{s}$ for $k_{s}$ large and integrate
term by term over $u=\cos \theta $, $\theta $ being the angle between the
vectors $\mathbf{p}_{2}$ and $\mathbf{k}$. When the conditions (\ref{analit}%
) hold, this operation is safe. Indeed, the expansion highlights factors 
\begin{eqnarray}
\frac{1}{(p_{1}\cdot k)^{3}} &=&\frac{1}{(E_{1}k_{0})^{3}}\sim \frac{1}{%
(E_{1}k_{s})^{3}},  \notag \\
\frac{1}{(p_{2}\cdot k)^{3}} &=&\frac{1}{(E_{2}k_{0}-\mathbf{p}_{2}\cdot 
\mathbf{k})^{3}}\sim \frac{1}{k_{s}^{3}(\pm E_{2}-p_{2s}u)^{3}},
\label{leading} \\
\frac{1}{[(p_{1}-p_{2})\cdot k]^{3}} &\mathbf{\sim }&\frac{1}{k_{s}^{3}(\pm
E_{1}\mp E_{2}+p_{2s}u)^{3}},  \notag
\end{eqnarray}%
where $p_{2s}=|\mathbf{p}_{2}|$. The subleading corrections have
denominators that are equal to powers of those shown in formula (\ref%
{leading}). We see that every term of the expansion leads to a regular $u$
integral. At the end, the logarithmic divergences are the coefficients of $%
\mathrm{d}k_{s}/k_{s}$.

The symmetry relation $I_{r,n,t}^{\text{nl\hspace{0.01in}d}}\left(
p_{1},p_{2}\right) =I_{t,n,r}^{\text{nl\hspace{0.01in}d}}\left(
p_{2},p_{1}\right) $ obviously holds, so we can assume, for example, $%
r\geqslant t$. By explicit calculation of $I_{r,n,t}$ for different values
of its indices, we find that 
\begin{equation}
I_{r,n,t}^{\text{nl\hspace{0.01in}d}}\left( p_{1},p_{2}\right) =0\qquad 
\text{unless }r+n+t\geqslant 4.  \label{che}
\end{equation}%
This result may be justified by evaluating the residues associated with the
propagators of $I_{r,n,t}$, as explained in subsection \ref{bubb}. The
residues with $k^{2}=$constant have a superficial degree of divergence $%
\omega _{0}=D-8+r+t$, while the residues with $\left( k+p_{1}\right) ^{2}=$%
constant and $\left( k+p_{2}\right) ^{2}=$constant have the generally larger
degrees $\omega _{1,2}=D-8+r+n+t$. Barring cancellations, the degree of
divergence of $I_{r,n,t}$ at $D=4$ is $\omega =r+n+t-4$, which is
nonnegative when $r+n+t\geqslant 4$.

Now we report the results of explicit calculations that confirm that $%
I_{r,n,t}$ is indeed nonlocally divergent when the inequality of (\ref{che})
holds. We begin with the massless case $m=0$. The simplest nontrivial
integral is 
\begin{equation}
I_{1,2,1}^{\text{nld}}\left( p_{1},p_{2}\right) =-i\frac{M^{8}}{128\pi ^{2}}%
\frac{1}{Q^{2}}\log \left( \frac{\Lambda _{\text{UV}}^{2}}{M^{2}}\right) ,
\label{i1}
\end{equation}%
where we have defined $Q^{2}\equiv -(p_{2}-p_{1})^{2}$. The term $1/Q^{2}$
is reminiscent of the pole-dominance models of form factors and seems to
signal --- if we insist with some physical interpretation --- the
propagation of a massless particle in the $t$-channel, with an ultraviolet
logarithmically divergent coefficient.

A more interesting case is provided by the amplitude 
\begin{equation}
I_{2,2,0}^{\text{nld}}\left( p_{1},p_{2}\right) =\frac{iM^{8}}{128\pi ^{2}}%
\left[ F(Q,p_{1},p_{2})-\frac{1}{Q^{2}}\right] \log \left( \frac{\Lambda
_{UV}^{2}}{M^{2}}\right) ,  \label{i2}
\end{equation}%
where%
\begin{equation*}
F(Q,p_{1},p_{2})=\frac{2}{\sqrt{\left( Q^{2}+p_{1}^{2}+p_{2}^{2}\right)
^{2}-4p_{1}^{2}p_{2}^{2}}}\text{arctanh}\frac{\sqrt{\left(
Q^{2}+p_{1}^{2}+p_{2}^{2}\right) ^{2}-4p_{1}^{2}p_{2}^{2}\,\,}}{%
Q^{2}+p_{1}^{2}-p_{2}^{2}}.
\end{equation*}%
In order to understand the dynamic properties of the first term of formula (%
\ref{i2}) (the second term has the same form as the one of the previous
amplitude), let us assume the kinematics of the Deep Inelastic Scattering,
i.e. 
\begin{equation*}
Q^{2}\,\gg \,\left\vert p_{1}^{2}\right\vert \,\approx \,\left\vert
p_{2}^{2}\right\vert \neq 0.
\end{equation*}%
Infrared singularities (soft and/or collinear) are then regulated by nonzero
virtualities of the external legs. Using the asymptotic expansion 
\begin{equation*}
\text{arctanh}(x)=-\frac{1}{2}\ln \left( \frac{1-x}{2}\right) +\mathcal{O}%
(1-x)\,\,\,\,\,\,\,\,\mathrm{for}\,\,\,x\lesssim 1,
\end{equation*}%
it is easy to show that 
\begin{equation}
F(Q,p_{1},p_{2})\simeq \frac{1}{Q^{2}}\ln \left( \frac{Q^{2}}{\mu _{2}^{2}}%
\right) ,  \label{fq}
\end{equation}%
where we have defined 
\begin{equation*}
\mu _{i}^{2}\equiv -p_{i}^{2}>\,0,\,\,\,\,\,i=1,2.
\end{equation*}%
It is convenient to assume $\mu _{2}^{2}>0$ in order to have a real
logarithm and then avoid absorptive parts related to the \textquotedblleft
decay\textquotedblright\ of the $p_{2}$ leg. Formula (\ref{fq}) exhibits a
collinear divergence for $\mu _{2}^{2}\rightarrow 0$ (or, equivalently, $%
Q^{2}\rightarrow +\infty $), which overlaps the nonlocal divergence already
found in the previous cases. The asymmetry of the result, namely the absence
of a collinear singularity for $\mu _{1}^{2}\rightarrow 0$, is related to
the fact that the power $(k\cdot p_{1})^{2}$ appearing in the numerator
screens the singularity related to the emission of a particle collinear to
the particle with momentum $p_{1}$. In the case of the previous amplitude $%
I_{1,2,1}$, collinear singularities related to the emission from any leg
were screened by the factor $(k\cdot p_{1})(k\cdot p_{2})$ in the numerator.
By generalizing the example just discussed, we expect that nonlocal
divergences overlap the usual, logarithmic infrared singularities found in
vertex functions.

It is natural to expect that nonlocal divergences also occur in one-loop box
diagrams, pentagon diagrams, etc., and that they overlap the usual
logarithmic structures (infrared divergences, small-$x$ logarithms, etc.).

We conclude by briefly reporting results concerning the massive case $m\neq
0 $. In both $I_{1,2,1}^{\text{nl\hspace{0.01in}d}}$ and $I_{2,2,0}^{\text{nl%
\hspace{0.01in}d}}$, it is sufficient to replace $M^{8}$ with $%
M^{10}/(M^{2}+im^{2})$ in formulas (\ref{i1}) and (\ref{i2}).

\section{Power counting for nonlocal divergences}

\setcounter{equation}{0}\label{pc}

The standard loop integrals in Minkowski spacetime are related to Euclidean
integrals by the Wick rotation, so the power counting rules governing their
ultraviolet behaviors are the same. This fact is actually far from trivial.
A Minkowski integral 
\begin{equation*}
\int \frac{\mathrm{d}^{D}k}{(2\pi )^{D}}\prod\nolimits_{i}\Delta
(k-p_{i},m_{i})
\end{equation*}%
is intuitively expected to be more singular than the corresponding Euclidean
integral in the ultraviolet region. Because of the Euclidean metric, the
Euclidean integrand falls off when any component of the momentum $k_{\mu }$
gets large. At first sight, the Minkowski pseudometric fails to provide an
equivalent suppression in several subdomains of integration, such as the
regions where the loop momentum is close to surfaces of the form $%
k^{2}+a\cdot k+b=0$ (determined by the poles of the propagators), where $a$
is a vector and $b$ is a constant. The reason why this intuitive argument is
not correct is that it overlooks the role played by the $+i\epsilon $
prescription, which allows the Wick rotation.

On the other hand, the higher-derivative propagator $S(k,m)$ has poles in
all quadrants, so the analytic continuation to Euclidean space is not
possible. Then the rules of power counting are no longer guaranteed to
coincide in Minkowski and Euclidean spaces and actually turn out to be
different.

To illustrate this fact, we work out a formula for the degree of divergence
of a generic one-loop diagram in Minkowski higher-derivative theories.
Assume that the propagators $S(k,m)$ behave like $1/(k^{2})^{N}$ for large $%
|k^{2}|$ and that the vertices contain up to $N^{\prime }$ derivatives.
Consider a one-particle irreducible diagram with $V$ vertices, equal to the
number of internal lines. We assume that $V>1$, i.e. exclude the tadpoles,
because they are independent of the external momenta and cannot originate
nonlocal divergences.

Letting $k$ denote the loop momentum, we integrate over the energy $k_{0}$
by means of the residue theorem. In each residue, $(k-q)^{2}$ is equal to
some constant, $q$ being a linear combination of external momenta. Making a
translation, we can assume that the integrand is evaluated at $k^{2}=$%
constant. Then by formula (\ref{partf}) the propagator $S(k,m)$ gives a
contribution that behaves like $1/k_{s}$ for large $k_{s}$, while each one
of the other $V-1$ propagators behaves like $1/((p-k)^{2})^{N}\sim 1/(p\cdot
k)^{N}$, where $p$ is also a linear combination of the external momenta. If
we use analyticity to assume $p^{2}>0$, the factors $1/(p\cdot k)^{N}$ are
regular everywhere.

On the other hand, the vertices provide at most $N^{\prime }V$ powers of $%
k_{s}$ and the integration measure is $k_{s}^{D-2}\mathrm{d}k_{s}$.
Collecting these pieces of information, the degree of divergence $\omega _{%
\text{nl}}$ of the $k_{s}$ integral is at most equal to%
\begin{equation}
\omega _{\text{nl}}=D-1+N^{\prime }V-1-(V-1)N=D-2+N+V(N^{\prime }-N).
\label{oml}
\end{equation}%
An integral with $\omega _{\text{nl}}<0$ is ultraviolet convergent, while an
integral with $\omega _{\text{nl}}\geqslant 0$ may be divergent. The
divergent parts are in general nonlocal, because, as shown in formula (\ref%
{leading}), the large $k_{s}$ expansion makes the ratios $1/(p\cdot k)^{N}$
factorize as $1/k_{s}^{N}$ times nonpolynomial functions of the external
momenta.

The condition $\omega _{\text{nl}}\geqslant 0$ is necessary to have a
divergence, but not sufficient. In many cases, it is possible to enhance it
by means of more sophisticated arguments. For example, it is possible to
show that the bubble diagram ($V=2$) benefits from an enhancement of one
unit when $N$ is odd. Then%
\begin{equation}
\omega _{\text{nl}}^{\prime }=D-3+2N^{\prime }-N.  \label{omla}
\end{equation}%
The reason is a simplification between the contributions $\sim 1/(p\cdot
k)^{N}$ of each propagator, calculated on the poles of the other propagator.

If $D=6$, $N=3$, $N^{\prime }=0$, $V=2$, which is the case treated in
subsection \ref{bubb}, we have $\omega _{\text{nl}}^{\prime }=0$, which
confirms that there is a logarithmic divergence. The same diagram in four
dimensions has no nonlocal divergences ($\omega _{\text{nl}}^{\prime }=-2$),
unless we equip it with nontrivial numerators. If we take $N^{\prime }=1$,
we raise $\omega _{\text{nl}}^{\prime }$ to 0, which is confirmed by the
nonvanishing coefficient $c_{0,1}$ of formula (\ref{cc}). On the other hand,
it is not enough to have a vertex with one derivative and a vertex with no
derivatives (which can be formally obtained by setting $N^{\prime }=1/2$),
as the vanishing of the coefficient $c_{1,0}$ confirms.

In the case of the triangle diagram ($D=4$, $N=3$ and $V=3$), we may
distribute the $r+2n+t$ derivatives over the three vertices by formally
writing $N^{\prime }=(r+2n+t)/3$. The integrals $I_{1,2,1}^{\text{nl\hspace{%
0.01in}d}}$ and $I_{2,2,0}^{\text{nl\hspace{0.01in}d}}$ have $N^{\prime }=2$
and $\omega _{\text{nl}}>0$, indeed formulas (\ref{i1}) and (\ref{i2}) shows
that they are divergent. Moreover, $r+2n+t<4$ implies $\omega _{\text{nl}}<0$%
, which agrees with formula (\ref{che}). A better agreement can be obtained
by improving the power counting as shown in section \ref{triangles}. Indeed,
after a residue is evaluated, a $k^{2}$ factor in the numerator does not
provide two powers of $k_{s}$, but one at most. This is equivalent to
setting $N^{\prime }=(r+n+t)/3$. Then formula (\ref{che}) follows in all
cases. Moreover, both $I_{1,2,1}^{\text{nl\hspace{0.01in}d}}$ and $%
I_{2,2,0}^{\text{nl\hspace{0.01in}d}}$ have $N^{\prime }=4/3$, $\omega _{%
\text{nl}}=0$, which implies that the ultraviolet divergence is at most
logarithmic, as is actually the case.

Let us inquire which theories have no nonlocal divergences at one loop, i.e.
when $\omega _{\text{nl}}<0$ for every $V>1$. Formula (\ref{oml}) shows that
this happens when $D-2<N-2N^{\prime }$ and $N\geqslant N^{\prime }$. All the
scalar and fermion theories with nonderivative interactions satisfy these
conditions for $N$ sufficiently large, in arbitrary dimensions. For example,
in four-dimensional scalar models with nonderivative interactions it is
sufficient to take $N=3$. As far as the fermions are concerned, assume that
their propagators $S_{F}(k,m)$ behave as $k^{\mu }/(k^{2})^{N}$ for large $%
|k^{2}|$. Then we can attach their numerator $k^{\mu }$ to a nearby vertex,
so the arguments given above apply with $N^{\prime }\rightarrow N^{\prime
}+1 $. The higher-derivative theories of gauge fields have $N^{\prime }=2N-1$%
, while those of gravity have $N^{\prime }=2N$, so neither of the two
satisfies the conditions for having no nonlocal divergences at one loop.
Both are expected to violate the locality of counterterms, if their
propagators have poles in the first or third quadrants. In the next section
we study the case of gravity explicitly.

\section{Higher-derivative gravity}

\setcounter{equation}{0} \label{QG}

In this section we use the results of the previous ones to work out the
nonlocal divergences of the graviton two-point function in a relatively
simple model of four-dimensional higher-derivative gravity with complex
poles. We simplify the calculations as much as possible by choosing a
specific Lagrangian and a convenient gauge fixing. The loop integrals are
linear combinations of the scalar integrals (\ref{imn}).

The simplest model of higher-derivative gravity is the Stelle theory \cite%
{stelle}, which contains the scalars $R$, $R^{2}$ and $R_{\mu \nu }R^{\mu
\nu }$. However, it is not suitable for our investigation, because its
propagators do not have poles in the first or third quadrants. The simplest
model with the features we need is the one with Lagrangian 
\begin{equation}
\mathcal{L}_{\text{HD}}=-\frac{\sqrt{-g}}{2\kappa ^{2}}\left[ R-\frac{1}{%
M^{4}}(D_{\rho }R_{\mu \nu })(D^{\rho }R^{\mu \nu })+\frac{1}{2M^{4}}%
(D_{\rho }R)(D^{\rho }R)\right] .  \label{lhd}
\end{equation}%
We expand the metric tensor $g_{\mu \nu }$ around the flat-space metric $%
\eta _{\mu \nu }=$diag$(1,-1,-1,-1)$ by writing%
\begin{equation*}
g_{\mu \nu }=\eta _{\mu \nu }+2\kappa h_{\mu \nu },
\end{equation*}%
where $\kappa $ is a constant of dimension $-1$ in units of mass and $h_{\mu
\nu }$ is the quantum fluctuation. After the expansion around flat space, we
raise and lower the indices by means of the flat-space metric. We further
define $h\equiv h_{\mu }^{\mu }$.

We choose the De Donder gauge-fixing function%
\begin{equation}
\mathcal{G}_{\mu }(g)=\eta ^{\nu \rho }\partial _{\rho }g_{\mu \nu }-\frac{1%
}{2}\eta ^{\nu \rho }\partial _{\mu }g_{\nu \rho }=\kappa (2\partial _{\nu
}h_{\mu }^{\nu }-\partial _{\mu }h)  \label{gmu}
\end{equation}%
and perform the gauge fixing as explained in appendix \ref{appe}. The
gauge-fixed Lagrangian then reads%
\begin{equation}
\mathcal{L}_{\text{grav}}=\mathcal{L}_{\text{HD}}+\frac{1}{4\kappa ^{2}}%
\mathcal{G}^{\mu }\left( 1+\frac{\square ^{2}}{M^{4}}\right) \mathcal{G}%
_{\mu }+\mathcal{L}_{\text{gh}},  \label{lgr}
\end{equation}%
where $\square =\eta ^{\mu \nu }\partial _{\mu }\partial _{\nu }$ is the
flat-space D'Alembertian, while the ghost Lagrangian is 
\begin{equation}
\mathcal{L}_{\text{gh}}=\bar{C}^{\mu }\left( 1+\frac{\square ^{2}}{M^{4}}%
\right) \left[ \square C_{\mu }-(2\delta _{\mu }^{\rho }\eta ^{\nu \sigma
}\partial _{\nu }-\eta ^{\rho \sigma }\partial _{\mu })\Gamma _{\rho \sigma
}^{\alpha }C_{\alpha }\right] .  \label{ghac}
\end{equation}%
The graviton propagator%
\begin{equation*}
\langle h_{\mu \nu }(p)\hspace{0.01in}\hspace{0.01in}h_{\rho \sigma
}(-p)\rangle _{0}=\frac{iM^{4}}{2(p^{2}+i\epsilon )}\frac{\eta _{\mu \rho
}\eta _{\nu \sigma }+\eta _{\mu \sigma }\eta _{\nu \rho }-\eta _{\mu \nu
}\eta _{\rho \sigma }}{(p^{2})^{2}+M^{4}}
\end{equation*}%
has the same form as that of the propagators of the previous sections, apart
from the constant matrices in the numerator.

Normally, the ghosts contribute to the renormalization, because they must
compensate the contributions of the temporal and longitudinal components of
the gauge fields, to give a total gauge invariant result. However, we can
easily show that in our case they can be ignored, because they cannot give
nonlocal divergences at one loop. Indeed, after the redefinition $\bar{C}%
^{\mu \prime }=\left( 1+\square ^{2}/M^{4}\right) \bar{C}^{\mu }$, the ghost
Lagrangian (\ref{ghac}) turns into the usual one, which is%
\begin{equation*}
\mathcal{L}_{\text{gh}}=\bar{C}^{\mu \hspace{0.01in}\prime }\left[ \square
C_{\mu }-(2\delta _{\mu }^{\rho }\eta ^{\nu \sigma }\partial _{\nu }-\eta
^{\rho \sigma }\partial _{\mu })\Gamma _{\rho \sigma }^{\alpha }C_{\alpha }%
\right] .
\end{equation*}%
For this reason, the ghost contribution to the graviton self-energy
coincides with the usual one, which has a local divergent part.

It is sufficient to work out the three-graviton vertex, since the one-loop
diagrams involving four-leg vertices are tadpoles, which can only have local
divergent parts. In the end, we just evaluate two diagrams, which are%
\begin{equation}
\includegraphics[width=12truecm]{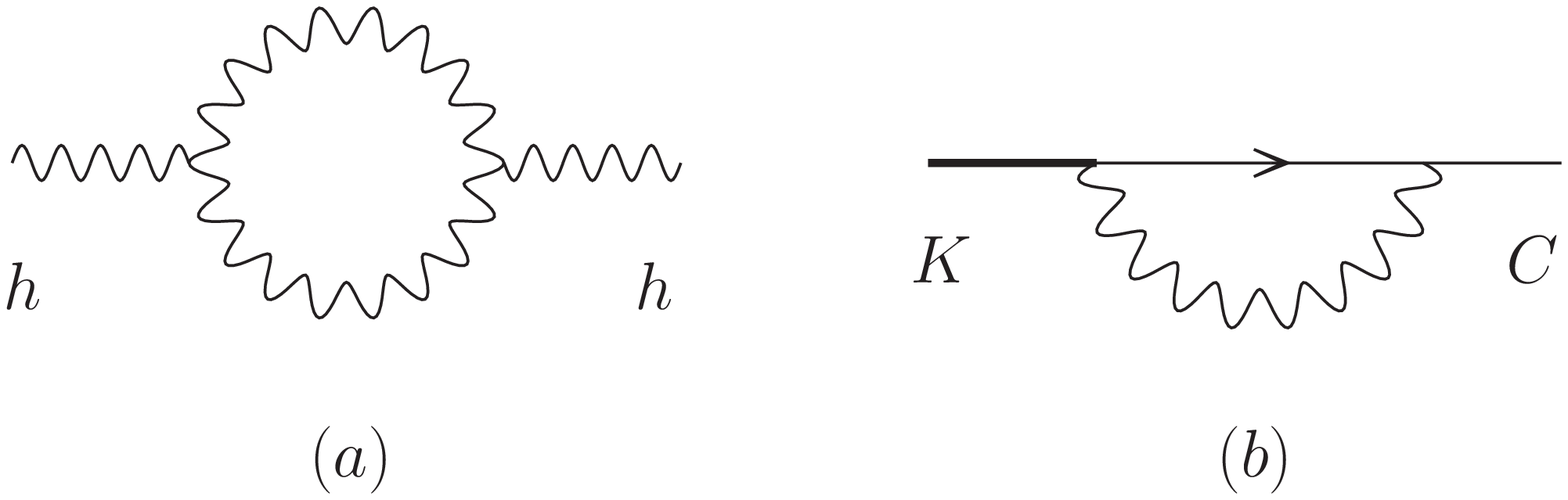}  \label{grafici}
\end{equation}%
where the wiggled line represents the graviton $h_{\mu \nu }$, the solid
line represents the source $K^{\mu \nu }$ coupled to the $h_{\mu \nu }$
transformation and the continuous line with the arrow represents the ghosts.
Diagram ($a$) encodes the nonlocal divergences of the graviton self-energy.
Diagram ($b$) encodes the nonlocal renormalization of the $h_{\mu \nu }$
transformation, which is necessary to derive the corrections to the Ward
identities satisfied by ($a$), as explained in appendix \ref{appe}.

\subsection{Results}

Given these ingredients, we are ready to perform the calculation, as well as
the consistency checks. We can reduce to the scalar integrals $I_{r,n}$ of
formula (\ref{imn}) by means of the Passarino--Veltman decomposition \cite%
{passarino}, which gives identities such as%
\begin{equation}
\mathcal{I}_{\mu _{1}\cdots \mu _{n}}(p)\equiv \int \frac{\mathrm{d}^{D}k}{%
(2\pi )^{D}}k^{\mu _{1}}\cdots k^{\mu _{n}}\hspace{0.01in}f(k^{2},p\cdot
k)=\sum_{i}A_{i}(p)\,T_{i}^{\mu _{1}\cdots \mu _{n}}(p),  \label{idat}
\end{equation}%
where $A_{i}(p)$ are scalar integrals and $T_{i}^{\mu _{1}\cdots \mu
_{n}}(p) $ are completely symmetric tensors built with $\eta ^{\mu \nu }$
and $p^{\mu }$. Since the graviton two-point functions has four indices, we
just need formula (\ref{idat}) for $n=1,2,3,4$. The cases $n=1,2$ give, for
example, 
\begin{eqnarray*}
\mathcal{I}_{\mu } &=&\frac{p^{\mu }}{p^{2}}\int \frac{\mathrm{d}^{D}k}{%
(2\pi )^{D}}(p\cdot k)\hspace{0.01in}f(k^{2},p\cdot k), \\
\mathcal{I}_{\mu \nu } &=&\frac{\eta ^{\mu \nu }}{3}\int \frac{\mathrm{d}%
^{D}k}{(2\pi )^{D}}\left[ k^{2}-\frac{(p\cdot k)^{2}}{p^{2}}\right]
f(k^{2},p\cdot k)-\frac{p^{\mu }p^{\nu }}{3p^{2}}\int \frac{\mathrm{d}^{D}k}{%
(2\pi )^{D}}\left[ k^{2}-4\frac{(p\cdot k)^{2}}{p^{2}}\right] f(k^{2},p\cdot
k).
\end{eqnarray*}

The one-loop nonlocal divergent part of the graviton two-point function is
equal to%
\begin{eqnarray}
\langle h_{\mu \nu }(p)\hspace{0.01in}\hspace{0.01in}h_{\rho \sigma
}(-p)\rangle _{1}^{\text{nl\hspace{0.01in}d}} &=&\frac{\kappa ^{2}M^{8}}{%
240\pi ^{2}(p^{2})^{2}}\left[ (68r+i)(\eta _{\mu \rho }\eta _{\nu \sigma
}+\eta _{\nu \rho }\eta _{\mu \sigma })+(373r-4i)\eta _{\mu \nu }\eta _{\rho
\sigma }\right.  \notag \\
&&-\frac{1}{8p^{2}}(125ir^{2}+544r+8i)\left( p_{\mu }p_{\rho }\eta _{\nu
\sigma }+p_{\mu }p_{\sigma }\eta _{\nu \rho }+p_{\nu }p_{\rho }\eta _{\mu
\sigma }+p_{\nu }p_{\sigma }\eta _{\mu \rho }\right)  \notag \\
&&+\frac{1}{4p^{2}}(255ir^{2}-1522r+36i)\left( p_{\mu }p_{\nu }\eta _{\rho
\sigma }+p_{\rho }p_{\sigma }\eta _{\mu \nu }\right)  \notag \\
&&\left. -\frac{1}{2(p^{2})^{2}}(185r^{3}+75ir^{2}-1048r+24i)p_{\mu }p_{\nu
}p_{\rho }p_{\sigma }\right] \ln \left( \frac{\Lambda _{UV}^{2}}{M^{2}}%
\right) ,  \label{res}
\end{eqnarray}%
where $r\equiv p^{2}/M^{2}$. It is easy to check that it is doubly
transverse, i.e.%
\begin{equation*}
p^{\nu }p^{\sigma }\langle h_{\mu \nu }(p)\hspace{0.01in}\hspace{0.01in}%
h_{\rho \sigma }(-p)\rangle _{1}^{\text{nl\hspace{0.01in}d}}=0
\end{equation*}%
up to local terms. However, it is not transverse, since $p^{\nu }\langle
h_{\mu \nu }(p)\hspace{0.01in}h_{\rho \sigma }(-p)\rangle _{1}^{\text{nl%
\hspace{0.01in}d}}$ does not vanish. The reason is that the gauge
transformation is itself affected by nonlocal divergences. The correct Ward
identity is (\ref{preW}), derived from diagram ($b$) as explained in
appendix \ref{appe}. It is easy to show that (\ref{res}) does satisfy (\ref%
{preW}), which provides a good check of the result.

Coherently with what we found in the previous sections, the divergences are
nonlocal and truly complex. It is impossible to subtract them away by means
of reparametrizations and (local as well as nonlocal) field redefinitions
that preserve Hermiticity.

In conclusion, Minkowski higher-derivative theories of gravity violate the
locality and Hermiticity of counterterms, when the propagators have poles in
the first or third quadrants. Gauge symmetries are unable to protect those
properties.

The gravitational Lagrangian (\ref{lhd}) is the simplest one that exhibits
the effects we have uncovered. Similar effects are expected to occur in the
theories with Lagrangians 
\begin{equation*}
\mathcal{L}_{\text{HD}}^{\prime }=-\frac{\sqrt{-g}}{2\kappa ^{2}}\left[ R+%
\frac{1}{M^{2}}R_{\mu \nu }P_{n}(\square _{c}/M^{2})R^{\mu \nu }-\frac{1}{%
2M^{2}}RQ_{n}(\square _{c}/M^{2})R\right] ,
\end{equation*}%
where $\square _{c}$ denotes the covariant D'Alembertian and $P_{n}$, $Q_{n}$
are real polynomials of degree $n>0$. The denominators of the free
propagators have the form $p^{2}R_{n+1}(p^{2})$, where $R_{n+1}$ is a real
polynomial of degree $n+1$. For every $n>0$, generic real polynomials $P_{n}$%
, $Q_{n}$ lead to poles in the first and third quadrants, which in turn
generate nonlocal, non-Hermitian divergences. Only by choosing the
polynomials $P_{n}$, $Q_{n}$ in very specific ways, we can manage to have
all the poles on the real axis. In that case, and only in that case, the
Euclidean and Minkowski theories are equivalent and the renormalization is
local.

If gravity is coupled to matter, we expect to find similar behaviors in the
matter sector. In particular, if the kinetic terms of the matter fields have
the same numbers of higher derivatives as the gravitational sector has, the
power counting is the same.

\section{Conclusions}

\label{concl} \setcounter{equation}{0}

We have shown that Minkowski higher-derivative quantum field theories whose
propagators have complex poles are generically inconsistent, because they
generate nonlocal, non-Hermitian ultraviolet divergences. Bubble diagrams,
for example, contain logarithmic divergences multiplied by inverse powers of
D'Alembertians. Triangle diagrams present more involved nonlocal
divergences, where ultraviolet effects mix with standard infrared effects.

Contrary to intuitive expectations, the introduction of higher-derivative
terms in the Lagrangian does not have a regulating effect, because the
constraints coming from power counting are much weaker. Indeed, the
contribution of one propagator calculated on the pole of another propagator
does not decay fast enough. This unusual behavior can also be explained by
the appearance of pinch singularities, unrelated to the usual absorptive
parts of amplitudes, which occur because the extra excitations introduced by
the higher derivatives come with effective prescriptions of both signs.

We have extended the calculations to higher-derivative quantum gravity and
proved, in particular, that gauge symmetries are unable to protect the
locality and Hermiticity of counterterms. The problems we have outlined add
up to the well-known problems that higher-derivative Minkowski theories have
with perturbative unitarity.

\vskip 12truept \noindent {\large \textbf{Acknowledgments}}

\vskip 12truept

One of us (D.A.) is grateful to M. Piva for useful discussions.

\appendix\renewcommand{\theequation}{\thesection.\arabic{equation}}

\section{Residue theorem in dimensional regularization}

\label{tricks}

\setcounter{equation}{0}

In this appendix we show that the dimensional regularization allows us to
evaluate the energy integrals in a straightforward way, even when they are
divergent: it is sufficient to sum the residues, while the contribution of
the integration path at infinity is always negligible.

The dimensional regularization is defined as follows. The integral on the
space momenta is continued to $D-1$ dimensions and done first. The energy
integral is not modified and done second. Strictly speaking, the two can be
exchanged when the energy integral is convergent. However, we show that it
is always legitimate to integrate on the energy first, if we apply the
residue theorem.

Without loss of generality, we can write a one-loop integral as a linear
combination of integrals of the form%
\begin{equation*}
\int_{-\infty }^{+\infty }\frac{\mathrm{d}E}{2\pi }\int \frac{\mathrm{d}%
^{D-1}\mathbf{k}}{(2\pi )^{D-1}}\frac{a(\mathbf{k})E^{s}}{%
E^{2r}+\sum_{i=1}^{2r}b_{i}(\mathbf{k})E^{2r-i}}
\end{equation*}%
where $a(\mathbf{k})$, $b_{i}(\mathbf{k})$ are polynomials of $\mathbf{k}$
and $r\geqslant 1$, $s\geqslant 0$ are integers. The denominators contain
the prescriptions to move the poles away from the real axis.

The energy integral is divergent if $s+1\geqslant 2r$, so we write $s=2r+n-1$
and take $n\geqslant 0$. When $|E|$ is much larger than all the other
scales, its divergent contributions are%
\begin{equation*}
\int_{|E|\sim \infty }\frac{\mathrm{d}E}{2\pi }\int \frac{\mathrm{d}^{D-1}%
\mathbf{k}}{(2\pi )^{D-1}}a(\mathbf{k})E^{n-1}\left( 1-\sum_{i=1}^{2r}\frac{%
b_{i}(\mathbf{k})}{E^{i}}-\sum_{i,j=1}^{2r}\frac{b_{i}(\mathbf{k})b_{j}(%
\mathbf{k})}{E^{i+j}}+\cdots \right) .
\end{equation*}%
All of the $\mathbf{k}$ integrals are integrals of polynomials of $\mathbf{k}
$ and give zero by the rules of the dimensional regularization. Thus, if we
close the integration path by means of a semicircle at infinity, in the
lower or upper half plane, we add a vanishing contribution. This allows us
to safely apply the residue theorem without having to worry about the
closure of the integration path.

\section{Gauge fixing and WTST\ identities in higher-derivative gravity}

\label{appe} \setcounter{equation}{0}

To handle the WTST identities \cite{WTST} of quantum gravity in a compact
form, we use the Batalin--Vilkovisky formalism \cite{bata}. We collect the
fields into the row%
\begin{equation*}
\Phi ^{\alpha }=\{h_{\mu \nu },C_{\mu },\bar{C}_{\mu },B_{\mu }\},
\end{equation*}%
where $C_{\mu }$, $\bar{C}_{\mu }$ and $B_{\mu }$ are the ghosts and the
antighosts of diffeomorphisms and the Lagrange multipliers for the gauge
fixing, respectively. We introduce conjugate sources 
\begin{equation*}
K_{\alpha }=\{K^{\mu \nu },K_{C}^{\mu },K_{\bar{C}}^{\mu },K_{B}^{\mu }\}
\end{equation*}%
and define the antiparentheses of two functionals $X$ and $Y$ of $\Phi $ and 
$K$ as 
\begin{equation*}
(X,Y)\equiv \int \left( \frac{\delta _{r}X}{\delta \Phi ^{\alpha }}\frac{%
\delta _{l}Y}{\delta K_{\alpha }}-\frac{\delta _{r}X}{\delta K_{\alpha }}%
\frac{\delta _{l}Y}{\delta \Phi ^{\alpha }}\right) ,
\end{equation*}%
where the integral is over the spacetime points associated with repeated
indices and the subscripts $l$, $r$ in $\delta _{l}$, $\delta _{r}$ denote
the left and right functional derivatives, respectively.

The total action is then%
\begin{equation*}
S(\Phi ,K)=S_{\text{HD}}+(S_{K},\Psi )+S_{K},
\end{equation*}%
where $S_{\text{HD}}=\int \mathcal{L}_{\text{HD}}$ is the classical action, $%
\Psi (\Phi )$ is a functional of the fields that performs the gauge fixing,
called gauge fermion, and the terms 
\begin{equation*}
S_{K}=-\int (\partial _{\mu }C_{\nu }+\partial _{\nu }C_{\mu }-2\Gamma _{\mu
\nu }^{\rho }C_{\rho })K^{\mu \nu }+\int g^{\nu \rho }C_{\rho }\left[
(\partial _{\mu }C_{\nu })+g^{\sigma \alpha }C_{\alpha }(\partial _{\sigma
}g_{\mu \nu })\right] K_{C}^{\mu }-\int B_{\mu }K_{\bar{C}}^{\mu }
\end{equation*}%
collect the symmetry transformations coupled to the external sources $K$.
The Lagrangian $\mathcal{L}_{\text{HD}}$ is given by formula (\ref{lhd}).

The action $S$ satisfies the master equation%
\begin{equation}
(S,S)=0,  \label{mast}
\end{equation}%
which collects the gauge invariance of $S_{\text{HD}}$ and the closure of
the symmetry transformations.

The generating functional $Z$ of the correlation functions and the
generating functional $W$ of the connected correlation functions are defined
by the formulas 
\begin{equation*}
Z(J,K)=\int [\mathrm{d}\Phi ]\exp \left( iS(\Phi ,K)+i\int \Phi ^{\alpha
}J_{\alpha }\right) =\exp iW(J,K),
\end{equation*}%
while the generating functional $\Gamma (\Phi ,K)=W(J,K)-\int \Phi ^{\alpha
}J_{\alpha }$ of the one-particle irreducible diagrams is the Legendre
transform of $W(J,K)$ with respect to $J$. Formula (\ref{mast}) implies that 
$\Gamma $ satisfies an identical master equation%
\begin{equation}
(\Gamma ,\Gamma )=0,  \label{mastg}
\end{equation}%
which collects the WTST identities in a compact form.

We choose the gauge fermion%
\begin{equation*}
\Psi =-\frac{1}{2}\int \bar{C}^{\mu }\left( 1+\frac{\square ^{2}}{M^{4}}%
\right) \left( \frac{1}{2}B_{\mu }-\frac{1}{\kappa }\mathcal{G}_{\mu
}\right) ,
\end{equation*}%
where $\mathcal{G}_{\mu }$ is given in formula (\ref{gmu}). Observe that the
indices of all the fields $\Phi ^{\alpha }$ and the sources $K_{\alpha }$
are raised and lowered by means of the flat-space metric. We find%
\begin{equation*}
(S_{K},\Psi )=-\frac{1}{4}\int B^{\mu }\left( 1+\frac{\square ^{2}}{M^{4}}%
\right) B_{\mu }+\frac{1}{2\kappa }\int B^{\mu }\left( 1+\frac{\square ^{2}}{%
M^{4}}\right) \mathcal{G}_{\mu }+S_{\text{gh}},
\end{equation*}%
where $S_{\text{gh}}=\int \mathcal{L}_{\text{gh}}$ and $\mathcal{L}_{\text{gh%
}}$ is given in formula (\ref{ghac}). We can integrate $B_{\mu }$ out, which
is equivalent to replacing it with the solution of its own field equation:%
\begin{equation*}
B_{\mu }=\frac{1}{\kappa }\mathcal{G}_{\mu }.
\end{equation*}%
So doing, we get%
\begin{equation*}
(S_{K},\Psi )\rightarrow \frac{1}{4\kappa ^{2}}\int \mathcal{G}^{\mu }\left(
1+\frac{\square ^{2}}{M^{4}}\right) \mathcal{G}_{\mu }+S_{\text{gh}}.
\end{equation*}%
The gauge-fixed Lagrangian is thus (\ref{lgr}).

Now we work out the WTST\ identity satisfied by the graviton two-point
function. Expand the functional $\Gamma $ as%
\begin{equation*}
\Gamma =S+\Gamma _{1}+\Gamma _{2}+\cdots ,
\end{equation*}%
where $\Gamma _{i}$ collects the contributions of the $i$-loop diagrams.
Note that $\Gamma _{i}$ cannot depend on $B$, $K_{\bar{C}}$ and $K_{B}$,
because no one-particle irreducible diagrams can be built with external legs
of this type. The master equation (\ref{mastg}) gives, at one loop,%
\begin{equation*}
(S,\Gamma _{1})=0.
\end{equation*}%
Expanding the antiparentheses on the left-hand side of this equation and
setting $\bar{C}=B=K=0$, we get%
\begin{equation}
\int \frac{\delta S_{\text{HD}}}{\delta h_{\mu \nu }}\left. \frac{\delta
_{l}\Gamma _{1}}{\delta K^{\mu \nu }}\right\vert _{\bar{C}=B=K=0}=\int \frac{%
\delta _{r}S_{K}}{\delta K^{\mu \nu }}\left. \frac{\delta _{l}\Gamma _{1}}{%
\delta h_{\mu \nu }}\right\vert _{\bar{C}=B=K=0}.  \label{masta}
\end{equation}%
This identity encodes the modified gauge invariance of the classical action $%
S_{\text{HD}}$. Indeed, it has the form%
\begin{equation}
\int \frac{\delta S_{\text{HD}}}{\delta h_{\mu \nu }}\Delta _{1}h_{\mu \nu
}+\int \frac{\delta \tilde{\Gamma}_{1}}{\delta h_{\mu \nu }}\Delta
_{0}h_{\mu \nu }=0,  \label{giv}
\end{equation}%
where $\tilde{\Gamma}_{1}=\left. \Gamma _{1}\right\vert _{\bar{C}=B=K=0}$
and 
\begin{equation*}
\Delta _{0}h_{\mu \nu }=-\frac{\delta _{r}S_{K}}{\delta K^{\mu \nu }}%
=\partial _{\mu }C_{\nu }+\partial _{\nu }C_{\mu }-2\Gamma _{\mu \nu }^{\rho
}C_{\rho },\qquad \Delta _{1}h_{\mu \nu }\left. =\frac{\delta _{l}\Gamma _{1}%
}{\delta K^{\mu \nu }}\right\vert _{\bar{C}=B=K=0}.
\end{equation*}%
In turn, the gauge invariance of $S_{\text{HD}}$, combined with formula (\ref%
{giv}), gives%
\begin{equation*}
\int \frac{\delta (S_{\text{HD}}+\tilde{\Gamma}_{1})}{\delta h_{\mu \nu }}%
(\Delta _{0}h_{\mu \nu }+\Delta _{1}h_{\mu \nu })=0,
\end{equation*}%
up to two-loop corrections. This identity states that the corrected action $%
S_{\text{HD}}+\tilde{\Gamma}_{1}$ is invariant under the corrected gauge
transformations $\Delta h_{\mu \nu }=\Delta _{0}h_{\mu \nu }+\Delta
_{1}h_{\mu \nu }$. The derivatives $\delta \tilde{\Gamma}_{1}/\delta h_{\mu
\nu }$ and $\left. \delta _{l}\Gamma _{1}/\delta K^{\mu \nu }\right\vert _{%
\bar{C}=B=K=0}$ are calculated through the diagrams ($a$) and ($b$) shown in
(\ref{grafici}), respectively.

Precisely, we can write%
\begin{eqnarray*}
\tilde{\Gamma}_{1} &=&-i\int h^{\mu \nu }(x)\langle h_{\mu \nu }(x)\hspace{%
0.01in}\rangle _{1}^{\text{1PI}}\hspace{0.01in}\mathrm{d}^{D}x-\frac{i}{2}%
\int h^{\mu \nu }(x)\langle h_{\mu \nu }(x)\hspace{0.01in}\hspace{0.01in}%
h_{\rho \sigma }(y)\rangle _{1}^{\text{1PI}}h^{\rho \sigma }(y)\hspace{0.01in%
}\mathrm{d}^{D}x\mathrm{d}^{D}y+\mathcal{O}(h^{3}), \\
\Delta _{1}h_{\mu \nu } &=&\left. \frac{\delta _{l}\Gamma _{1}}{\delta
K^{\mu \nu }}\right\vert _{\bar{C}=B=K=0}=\left. \frac{\delta _{l}W_{1}}{%
\delta K^{\mu \nu }}\right\vert _{\bar{C}=B=K=0}=\langle \partial _{\mu
}C_{\nu }+\partial _{\nu }C_{\mu }\rangle _{1,J}^{\text{1PI}}-2\langle
\Gamma _{\mu \nu }^{\rho }C_{\rho }\rangle _{1,J}^{\text{1PI}}+\mathcal{O}%
(Ch),
\end{eqnarray*}%
where the one-loop correlation functions $\langle \cdots \hspace{0.01in}%
\rangle _{1,J}$ are evaluated at nonvanishing external sources $J$.

Note that $\langle h_{\mu \nu }\hspace{0.01in}\rangle _{1}^{\text{1PI}}$ has
no nonlocal divergences, because it is a tadpole. On the other hand, $%
\langle \partial _{\mu }C_{\nu }\hspace{0.01in}\rangle _{1,J}^{\text{1PI}}=0$%
, because the insertion is linear in the fields. As far as the term $%
-2\langle \Gamma _{\mu \nu }^{\rho }C_{\rho }\rangle _{1,J}^{\text{1PI}}$ is
concerned, we are just interested in its nonlocal divergent part to the
zeroth order in $h_{\mu \nu }$, which we denote by $\left. \Delta _{1}h_{\mu
\nu }\right\vert _{h=0}^{\text{nl\hspace{0.01in}d}}$. We get, in momentum
space,%
\begin{eqnarray*}
&&\left. \Delta _{1}h_{\mu \nu }\right\vert _{h=0}^{\text{nl\hspace{0.01in}d 
}}(p)=-\frac{\kappa ^{2}M^{8}}{96\pi ^{2}(p^{2})^{4}}\left[
(3ir^{2}+2r-2i)p^{2}(p_{\mu }\delta _{\nu }^{\rho }+p_{\nu }\delta _{\mu
}^{\rho })\right. \\
&&\qquad \qquad +\left. p^{2}(3r-2i)p^{\rho }\eta _{\mu \nu
}+4(-4r+3i)p^{\rho }p_{\mu }p_{\nu }\right] C_{\rho }(p)\ln \left( \frac{%
\Lambda _{UV}^{2}}{M^{2}}\right) ,
\end{eqnarray*}%
where $r=p^{2}/M^{2}$. The quadratic part of $S_{\text{HD}}$ is, in momentum
space,%
\begin{equation*}
\frac{1}{4}\int h^{\mu \nu }(-p)\left[ 1+\frac{(p^{2})^{2}}{M^{4}}\right]
p^{2}\left( \Pi _{\mu \rho }\Pi _{\nu \sigma }+\Pi _{\mu \sigma }\Pi _{\nu
\rho }-2\Pi _{\mu \nu }\Pi _{\rho \sigma }\right) h^{\rho \sigma }(p),
\end{equation*}%
where $\Pi _{\mu \nu }=\eta _{\mu \nu }-p_{\mu }p_{\nu }/p^{2}$. Inserting
these expressions in (\ref{masta}), we find the Ward identity 
\begin{equation}
p^{\nu }\langle h_{\mu \nu }(p)\hspace{0.01in}h_{\rho \sigma }(-p)\rangle
_{1}^{\text{nl\hspace{0.01in}d}}=-\left( \eta _{\rho \sigma }p^{2}-p_{\rho
}p_{\sigma }\right) p_{\mu }\frac{\kappa ^{2}M^{8}}{96\pi ^{2}(p^{2})^{3}}%
\left( 1+r^{2}\right) (3r-2i)\ln \left( \frac{\Lambda _{UV}^{2}}{M^{2}}%
\right) ,  \label{preW}
\end{equation}%
up to local corrections.

\end{document}